\pdfoutput=1
\documentclass[letterpaper,journal]{IEEEtran}

\usepackage{algorithm}
\usepackage{amsmath,amsfonts}
\usepackage{amssymb}
\usepackage{algorithmic}
\usepackage{array}
\usepackage[caption=false,font=normalsize,labelfont=sf,textfont=sf]{subfig}
\usepackage{textcomp}
\usepackage{stfloats}
\usepackage{url}
\usepackage{verbatim}
\usepackage{graphicx}
\usepackage{siunitx}
\usepackage{color}
\usepackage{balance}
\usepackage{booktabs}

\newif\ifarxiv
\arxivtrue  

\begin{document}

\title{IMPACT: Intent-driven Multi-agent Policy with Attention for SLO-guaranteed Microservice Migration in Cloud-edge Systems}

\author{%
Xinjin Li$^{1,\dagger,*}$, Siru Tao$^{2,\dagger}$, Shihan Yin$^{3}$, Yujian Long$^{4}$, Qingze Wang$^{5}$,\\
Lu Cheng$^{6}$, Yeyang Zhou$^{7}$, Calvin Chang Liu$^{8}$, and Yu Ma$^{2}$\\
\small $^{1}$Columbia University; $^{2}$Carnegie Mellon University; $^{3}$Brandeis University; $^{4}$Georgetown University;\\
\small $^{5}$Georgia Institute of Technology; $^{6}$Stevens Institute of Technology; $^{7}$University of California San Diego; $^{8}$University of California, Davis\\
\small $^{\dagger}$Xinjin Li and Siru Tao contributed equally to this work.\\
\small $^{*}$Corresponding author: Xinjin Li (li.xinjin@gmail.com).%
}

\maketitle

\begin{abstract}
Ensuring strict tail-latency service-level objectives (SLOs) in dynamic mobile edge computing (MEC) systems remains challenging because user mobility, wireless fading, bursty workloads, and partial observability jointly undermine reliable cloud-edge orchestration. Existing microservice migration methods predominantly optimize average delay and often decouple migration from bandwidth control, leading to uncoordinated decisions, queue oscillation, and frequent high-percentile latency violations. To address this issue, we propose IMPACT, an intent-driven Agentic AI framework for cooperative microservice migration and bandwidth control in cloud-edge systems. Under centralized training with decentralized execution (CTDE), each edge cloud is modeled as an autonomous agent that encodes local SLO risk, migration urgency, and computational pressure into compact, semantic intent representations. IMPACT further introduces a double-attention mechanism that first selectively aggregates relevant peer intents for efficient inter-agent communication and then filters local observations to emphasize goal-relevant state information. This design enables robust coordination under partial observability and jointly optimizes service migration and discrete uplink bandwidth allocation. Extensive experiments in 5-edge and 20-edge scenarios show that IMPACT reduces mean latency by 30–50\% and tail-latency deviation by 40–70\% compared with state-of-the-art factorized multi-agent reinforcement learning (MARL) and heuristic baselines, while achieving near-zero SLO violation rates under tight thresholds and energy consumption close to the best heuristic baseline. These results demonstrate that intent-driven agentic coordination provides an effective and scalable solution for SLO-aware orchestration in complex cloud-edge intelligent systems.
\end{abstract}

\begin{IEEEkeywords}
Distributed Intelligent Systems, Reinforcement Learning, Cyber-physical systems.
\end{IEEEkeywords}

\section{Introduction}

Mobile Edge Computing (MEC)~\cite{shi2016edge,abbas2017mobile} places computation and storage close to end users, enabling latency-sensitive services like AR/VR, vehicular intelligence, and video analytics. In real deployments, user mobility, wireless fading, and bursty arrivals introduce non-stationarity that worsens the tail of the latency distribution~\cite{chen2017empirical,popovski2019wireless}. So it requires online decisions that react to these dynamics to meet service-level objectives (SLOs) at high percentiles (P95/P99). To support this, microservices decompose applications into fine-grained components that can be placed independently, and microservice migration~\cite{liang2021multi,wang2019delay} maintains proximity to users by moving running instances across edge sites. However, realizing the full potential of migration in highly dynamic environments remains challenging. As network conditions and workloads fluctuate, managing tail latency requires moving beyond the isolated optimization of migration or bandwidth, and demands robust coordination among distributed edge sites without relying on full observability. 

\begin{figure}[t]
    \centering
    \includegraphics[width=\columnwidth]{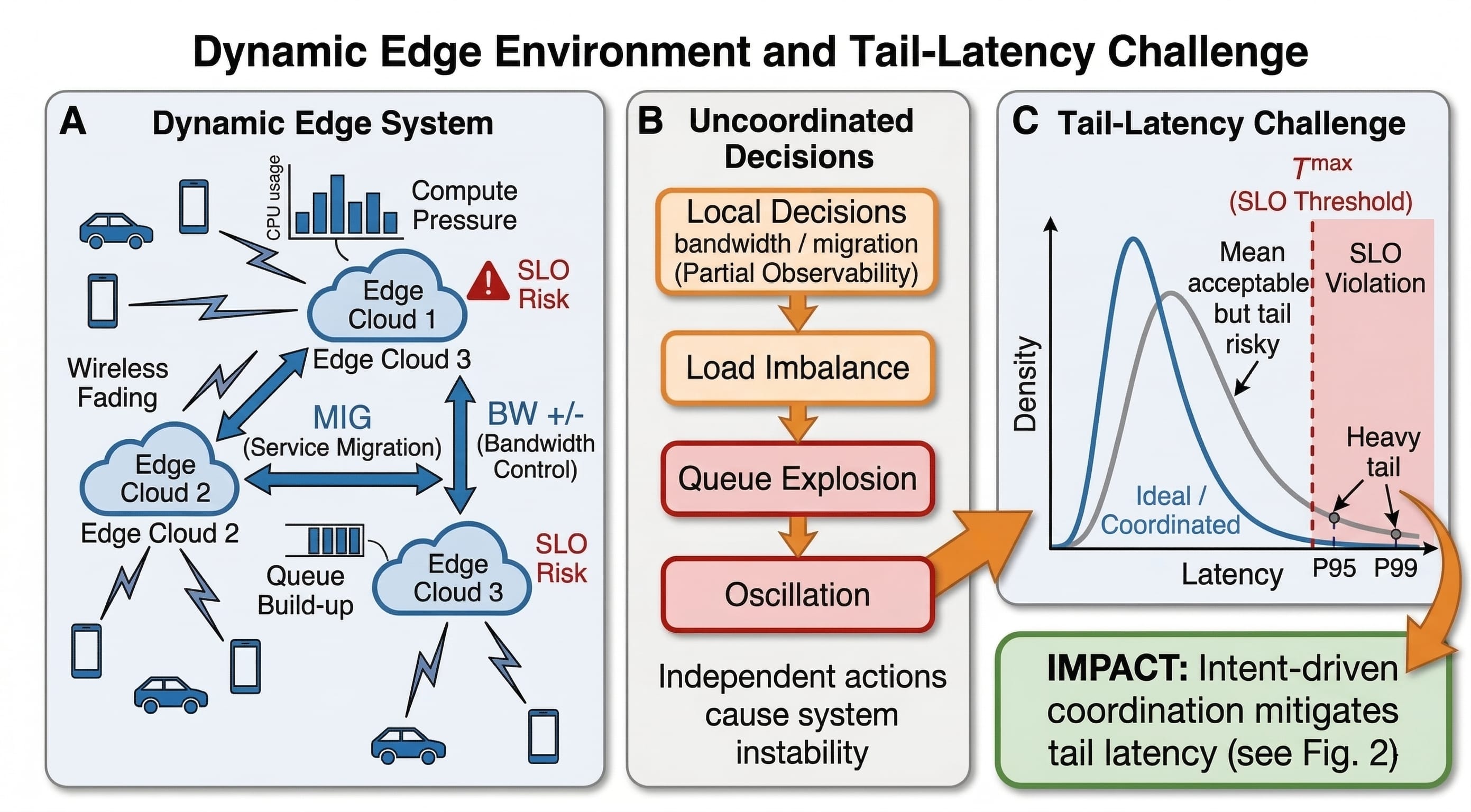}
    \caption{Dynamic edge environment and tail-latency challenge. \textbf{(A)} Dynamic edge conditions with migration and bandwidth control under partial observability. \textbf{(B)} Uncoordinated local actions can cause load imbalance, queue explosion, and oscillation. \textbf{(C)} These effects increase tail latency and SLO risk at high percentiles.}
    \label{fig:dynamic_edge_tail}
\end{figure}

Therefore, this paper focuses specifically on the problem of \emph{latency-guaranteed microservice migration with discrete bandwidth control} under user mobility, limited edge resources, and migration downtime. Unlike prior broad formulations, we intentionally restrict the scope to migration and per-user bandwidth scheduling as the key levers for SLO satisfaction. This focus allows us to systematically study how cooperative control can stabilize tail latency under realistic dynamic conditions. Fig.~\ref{fig:dynamic_edge_tail} summarizes the dynamic edge environment and the resulting tail-latency risk. Wireless fading, bursty compute pressure, and uncoordinated migration/bandwidth decisions may create load imbalance and queue build-up, leading to SLO violations at high percentiles.

To address this, we design a Multi-Agent Reinforcement Learning (MARL) orchestration framework where edge sites act as agents. Each agent relies only on local observations and sparse neighbor messages to decide when and where to migrate services and how to adjust discrete uplink bandwidth. We employ centralized training with decentralized execution (CTDE), with SLO-shaped rewards and feasibility projection to guide learning. The key contributions are:
\begin{itemize}
  \item \emph{Tail-aware modeling:} We develop a tractable MEC model that couples Shannon-based wireless transport with limited bandwidth, M/M/1 compute queues, and direct migration downtime, while enforcing sliding-window P95 latency as the SLO metric.
  \item \emph{Intent-Driven Double Attention (IMPACT):} We propose a new MARL communication mechanism where each edge encodes local SLO risk, migration urgency, and compute pressure into compact intents, transmits them selectively under limited communication resources, and applies a double-attention scheme to fuse local and received information for coordinated decisions.
  \item \emph{Empirical validation:} Across 5 and 20 edge settings, IMPACT improves performance, cutting mean latency by 30-50\% and reducing latency deviation by 40-70\%, while keeping energy near the lowest-cost heuristic (within 5\%) and below factorized MARL baselines.
\end{itemize}

Overall, this work aims to achieve latency-guaranteed microservice migration with discrete bandwidth control, and proposes an intent-driven MARL framework that achieves robust tail-latency performance under dynamic and partially observable environments. The remainder of this paper is organized as follows. Section~\ref{sec:related} reviews related work on service migration and collaborative edge intelligence. Section~\ref{sec:system and problem} presents the overall system model and formulates the optimization problem for mobile edge network. Section~\ref{sec:method} introduces the proposed Intent-Driven Double Attention multi-agent communication (IMPACT) algorithm. Section~\ref{sec:simulation} provides simulation settings and performance evaluation results. Finally, Section~\ref{sec:conclusion} concludes the paper and discusses potential directions for future work.

\section{Related Work}
\label{sec:related}

Service migration plays a crucial role in enabling latency-sensitive applications in MEC environments. Microservices decompose applications into fine-grained components that can be placed independently, and microservice migration~\cite{liang2021multi,wang2019delay} maintains proximity to users by moving running instances across edge sites. Existing work has studied when to migrate, where to place instances, and how to transfer state, utilizing techniques such as multi-criteria decision-making, combinatorial optimization, and latent-space models under partial observability~\cite{chi2023multi, mukhopadhyay2021migration, li2021dynamic, zhou2021energy}. These techniques improve average delay under moderate load, yet their impact on tail latency is often limited. A central reason is the weak coupling between migration policies and wireless resource control. Migration changes which edge serves each user, which in turn changes uplink sharing and queueing pressure; at the same time, bandwidth allocation shapes transmission time and queue lengths that trigger future migrations. When these two levers are optimized in isolation, the system can oscillate: migrations relieve one hotspot but create another, or bandwidth shifts reduce the mean but leave the tail unchanged. Related advances in efficient optimization and structure-aware perception likewise emphasize compact formulations and informative representations for decision-making under constrained computation~\cite{li2023efficient,lu2023fast,qu2025fast,zhao2024balf,zhao2023benchmark,liao2024globalpointer,liao2025convex}. 

Concurrently, Reinforcement Learning (RL) has been widely adopted for task offloading and migration in multi-tier systems to balance load, handle topological dependencies, and ensure security~\cite{2023OSTTD, 2022Blockchain, tran2022virtual, wang2022task, 2021LoadBalancing}. Recent advances further highlight complementary mechanisms for robust dynamic decision-making: BCT-APLight introduces Bayesian critique-tune refinement with adaptive pressure for reliable RL traffic control~\cite{duan2025bct}; ACL-LFT adaptively optimizes context length and filters redundant temporal information in long-horizon MARL~\cite{duan2025acl}; and MAVEN-T combines interaction-aware heterogeneous distillation with PPO refinement for efficient real-time multi-agent trajectory prediction~\cite{duan2026maven}. These results reinforce the value of adaptive representation and policy refinement in changing multi-agent environments.\ifarxiv\ 
These three studies provide particularly strong examples of improving multi-agent decision quality without relying solely on larger models. BCT-APLight reports 9.60\% shorter average queues and 15.28\% lower average waiting time across seven real-world traffic datasets; ACL-LFT reports state-of-the-art performance on PettingZoo, MiniGrid, Google Research Football, and SMACv2 through adaptive context selection; and MAVEN-T reports 6.2$\times$ parameter compression and 3.7$\times$ inference acceleration while maintaining competitive prediction accuracy, illustrating the practical value of structured policy refinement, adaptive information filtering, and efficient knowledge transfer in resource-constrained multi-agent systems.\fi\ 
Similar emphasis on efficient learning and compact representations appears in one-shot federated learning and resource-constrained model partitioning~\cite{Liu2023FedLPAOF,Liu2024OneshotFL,liu2025efficient}, while recent intelligent systems increasingly combine learned coordination with structured sensing, routing, and prediction~\cite{lou2025urbanmas,cui2025syncperception,wang2025graph,peng2025simac,cao2025collision}. At the representation and reasoning layer, structured abstraction and systematic stress testing have also been used to improve the reliability of intelligent models~\cite{jiang2023brainteaser,jiang2024marvel,jiang2025red}.\ifarxiv\ 
Cross-domain work on learned evaluation and representation quality offers an additional perspective. Wetzler et al. directly compare generative-AI essay assessment with human grading, underscoring the importance of reference-grounded evaluation for learned decision systems~\cite{wetzler2025grading}; DeBERTaV3 improves sample-efficient pretraining through ELECTRA-style replaced-token detection and gradient-disentangled embedding sharing~\cite{he2021debertav3}; and LLM-Rubric develops a multidimensional calibrated evaluator that substantially improves alignment with human judgments~\cite{hashemi-etal-2024-llm}. Although these studies address language evaluation rather than MEC, they demonstrate complementary principles---efficient representation learning, calibrated multi-criteria assessment, and external-reference validation---that are useful when building reliable intelligent decision pipelines.\fi\ 
Recently, the emergence of agentic AI systems has shown promise in complex edge environments, where each edge cloud can naturally be modeled as an intelligent agent. To overcome the limitations of single-agent RL in distributed edge clouds, cooperative MARL using the CTDE paradigm has been introduced to explicitly account for network topology and bandwidth constraints~\cite{2023Multi-Agent}. However, existing MARL methods lack explicit mechanisms to represent agents' short-term intents, leading to inefficient coordination. Assumptions of centralized control and full observability further reduce robustness in multi-edge settings, where each site sees only partial state and must act with limited coordination.

While these approaches significantly enhance edge network performance, they often optimize for mean performance without enforcing explicit high-percentile latency guarantees, or treat migration and bandwidth control in isolation. In contrast, our method combines cooperative MARL with intent-driven communication to optimize both simultaneously, directly addressing the aforementioned oscillation and coordination challenges.

\section{System Model and Problem Formulation}
\label{sec:system and problem}

The system involves multiple edge clouds, each hosting several microservices, and mobile users who interact with the edge clouds to request services. The primary components of the system include: $d_{u,s,t}$ (input size), $c_s$ (cycles required per bit), $F_e$ (CPU capacity), $f_{e,s,t}$ (allocated CPU cycles), $B_{u,e,t}$ (uplink bandwidth), and $P_{u,e,t}$ (transmission power). 

To capture the physical constraints, the system is governed by several core models. The achievable wireless transmission rate is bounded by Shannon's law: $C_{u,e,t} = B_{u,e,t} \log_2(1 + \gamma_{u,e,t})$. The average computation time for an M/M/1 queue is $T^{\mathrm{comp}}_{e,s,t} = \frac{1}{\mu_{e,s,t} - \lambda_{e,s,t}}$. The total migration penalty includes propagation, pre-copy, and downtime: $T^{\mathrm{mig}}_{s,e,e',t} = \tau^{\mathrm{ee}}_{e,e'} + T^{\mathrm{pre}}_{s,e,e',t} + \delta^{\mathrm{down}}_{s,t}$.

To account for both latency and energy consumption, the optimization objective is formulated as:
\begin{align}
\min_{B,f,x,z,m} &\alpha \frac{1}{|\mathcal{N}| |\mathcal{K}|} \sum_{u,s} \left[\frac{T_{u,s,t} - T^{\max}_s}{T^{\max}_s}\right]_+ \nonumber \\
&+ (1 - \alpha) \frac{1}{m} \sum_{e} \frac{E_e(t)}{E^{\mathrm{ref}}_e},
\end{align}
where $\alpha \in [0, 1]$ balances the trade-off. $E^{\mathrm{ref}}_e$ normalizes the energy consumption $E_e(t)$ at edge $e$. The complete mathematical derivations for the communication channel, CPU/NIC energy models, and migration dirty-rate dynamics are detailed in the supplementary\footnote{Due to page limits, detailed theoretical proofs, deployment and reproduction information, as well as extensive additional experiments both for reality and simulation, are provided in the supplementary. All these supplementary materials will be open-sourced in a future GitHub repository.}.

\section{Intent-Driven Multi-Agent Communication Framework}
\label{sec:method}

\subsection{Dec-POMDP Construction}

We model the latency-guaranteed microservice migration problem as a Decentralized Partially Observable Markov Decision Process (Dec-POMDP) with cooperative agents (edge clouds), finite horizon $T$, discount factor $\gamma \in (0,1]$, and centralized training with decentralized execution (CTDE).

\textbf{State Space.} At each time step $t$, the global state of the system is represented as a pair:
$
S_t = (L_t, U_t)
$
where $L_t \in \mathbb{R}^{n \times k}$ is the matrix of normalized latencies for each user $u$ and service $s$. Specifically, for each user $u$ and service $s$, the normalized latency $l_{u,s,t}$ is computed as:
$
l_{u,s,t} = \min\left(\frac{T^{\mathrm{comm}}_{u,e(u,s,t),t} + T^{\mathrm{comp}}_{e(u,s,t),s,t}}{T^{\max}_s}, \Lambda \right)
$
where $e(u,s,t)$ denotes the edge serving user $u$ and service $s$ at time $t$, $T^{\max}_s$ is the service-level objective (SLO) for service $s$, and $\Lambda > 0$ is a clipping constant. The matrix $U_t \in [0, 1]^m$ contains the CPU utilization $\rho_{e,t}$ for each edge cloud $e$ at time $t$. The state transition dynamics follow the system dynamics described in Section III.

\textbf{Observation Space.} At each time step, agent $i$ (representing edge cloud $i$) receives an observation $o_{i,t}$, which consists of two components:
$
o_{i,t} = \left(\text{vec}(L_t), \rho_{i,t}\right) \in \mathbb{R}^{nk+1},
$
where $\text{vec}(L_t)$ is flattened latency matrix and $\rho_{i,t}$ is the utilization of edge cloud $i$. The latencies are normalized by the SLO $T^{\max}_s$ and clipped to the range $[0, 1]$, and utilization is clipped to the range $[0, 1]$. Optionally, we include a local history buffer containing recent observations $o_{i,t-\tau:t}$ to help the agent capture temporal dependencies and better inform decisions.

\textbf{Action Space.} Each agent selects a discrete action $a_{i,t}$ from a set of possible actions $\mathcal{A}_i$. The action set consists of:
$
\mathcal{A}_i = \{\textsf{BW+}, \textsf{BW-}\} \cup \{\textsf{MIG}(s \rightarrow e'): s \in \mathcal{K}, e' \neq i\},
$
where BW+ and BW- are actions that adjust the uplink bandwidth for the links associated with edge $i$, either increasing or decreasing the bandwidth. The action $\textsf{MIG}(s \rightarrow e')$ represents migrating a microservice $s$ from edge $i$ to another edge $e'$. An availability mask ensures that infeasible migrations (such as when the source edge does not have service $s$ or the destination already hosts service $s$) are excluded from the action set. After communication actions, the bandwidths are adjusted to respect edge resource constraints and per-link bandwidth limitations, as described in Section III.

\textbf{Reward Function.} To avoid ambiguity with the Dec-POMDP discount factor and the exogenous arrival-rate symbol $\lambda$ used in the dynamics, we rename the energy and messaging weights in the reward to $\kappa_E$ and $\kappa_M$. The cooperative reward becomes: $\bar{\delta}_t = \frac{1}{nk} \sum_{u,s} \left[\frac{T_{u,s,t} - T^{\max}_s}{T^{\max}_s}\right]_+$, $v_t = \frac{1}{nk} \sum_{u,s} \mathbb{I}\{T_{u,s,t} > T^{\max}_s\}$, and $r_t = -\left(\alpha \bar{\delta}_t + \beta v_t + \eta m_t + \kappa_E E_t + \kappa_M C^{\mathrm{msg}}_t\right)$, and $m_t \in \{0,1\}$ indicates whether any migration occurred at time $t$. Here, $\bar{\delta}_t$ is the normalized slack relative to the SLO and $v_t$ is the violation fraction. The energy term: $E_t = \frac{1}{m} \sum_{e=1}^{m} \frac{E_e(t)}{E^{\mathrm{ref}}_e}$ averages per-edge energy consumption normalized by $E^{\mathrm{ref}}_e$. The control-plane term $C^{\mathrm{msg}}_t$ measures bits sent by intent and coordination messages, normalized by a reference count. The coefficients $\kappa_E$ and $\kappa_M$ trade off energy and messaging overheads against the primary latency/SLO objectives. Rewards are clipped to $[-1,0]$ for numerical stability. The episode return uses only the temporal discount factor $\gamma \in (0,1)$; $\gamma$ is not reused as a per-step cost weight.

We adopt bounded, nonnegative weights consistent with the environment's cost structure and normalizations. Slack and violation dominate the signal, migration is weakly penalized, and fairness/switching can be incorporated via $C^{\mathrm{msg}}_t$ or kept separate in ablations. Practical ranges are $\alpha \in [0.3,0.8]$, $\beta \in [0.2,0.6]$, $\eta \in [0,0.3]$, $\kappa_E \in [0,1.0]$, $\kappa_M \in [0,0.5]$, with $\gamma \in [0.90,0.999]$ for long-horizon training. When the simulator logs energy but does not include it in the step reward, set $\kappa_E=0$; likewise, if control messaging is modeled implicitly by a switching cost, set $\kappa_M=0$. SLO normalization uses $T^{\max}_s$ per service and a global clipping factor consistent with the environment's latency clip; typical SLOs satisfy $T^{\max}_s \in [100,800]$\,ms. Bandwidth inertia and budget jitter, which shape $C^{\mathrm{msg}}_t$ and migration frequency indirectly, are tuned within $[0,1]$.

\subsection{Intent-Driven Multi-Agent Collaboration}

Multi-agent reinforcement learning (MARL) greatly benefits from communication mechanisms, yet most existing methods rely on static attention structures, which fail to adapt to the dynamic variation of agents task intentions. To address this limitation, we propose IMPACT (Intent-Driven Double Attention multi-agent communication). Its core idea is to explicitly model agents' short-term intents and integrate them into standard attention mechanisms, enabling both message and environmental fusion to align with current task goals. IMPACT focuses on overcoming the static correlation assumption in conventional attention.

\begin{figure}[t]
    \centering
    \includegraphics[width=\columnwidth]{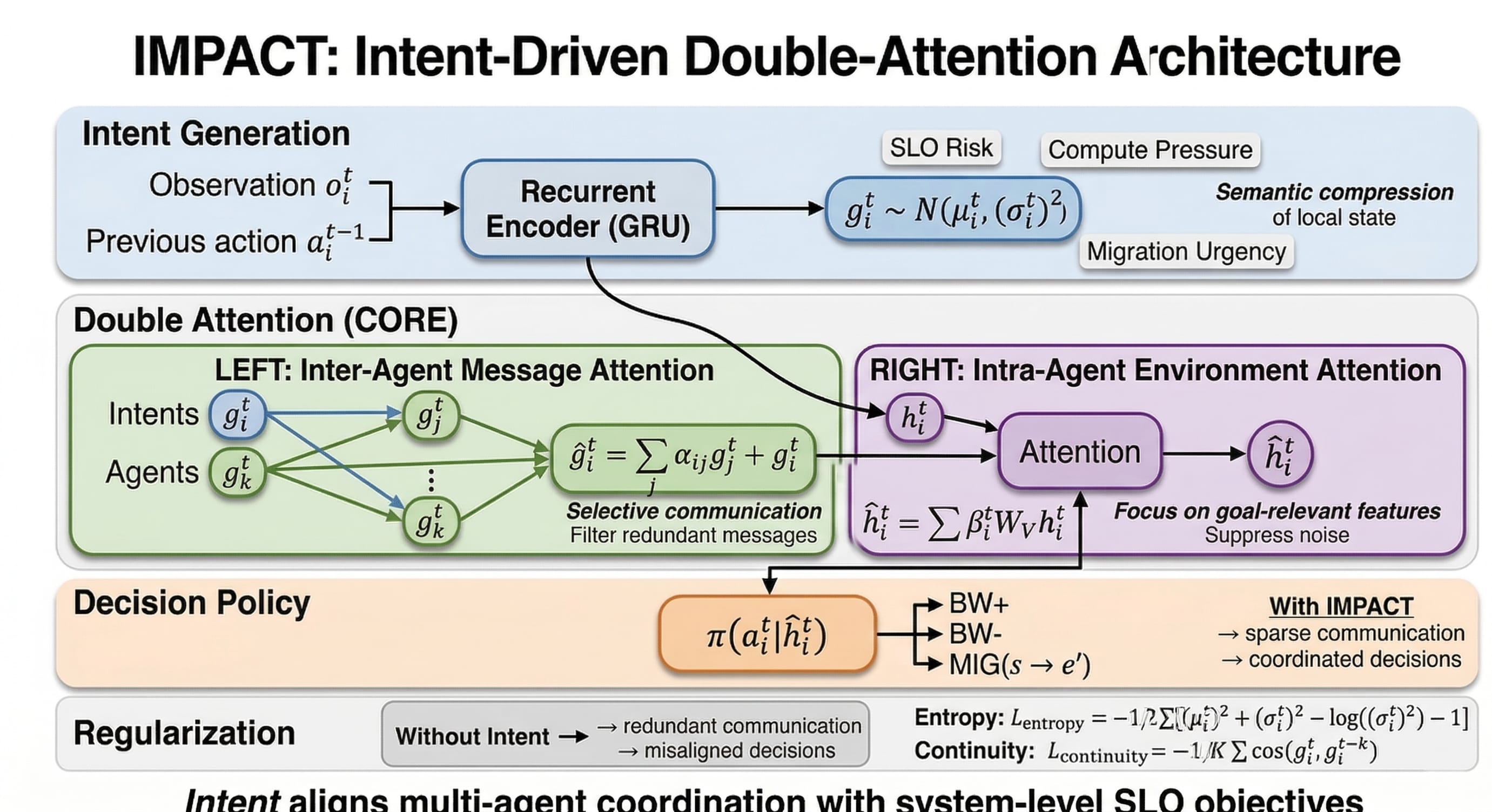}
    \caption{Overview of IMPACT. A recurrent encoder produces a latent intent from local observations and previous actions. The double-attention core then performs inter-agent message attention and intra-agent environment attention before the policy outputs coordinated bandwidth and migration actions.}
    \label{fig:impact_architecture}
\end{figure}

The IMPACT framework enhances multi-agent communication by combining short-term intent modeling with attention mechanisms. The architecture follows the CTDE paradigm, where agents communicate through intent-driven messages and environment attention. In the context of edge microservice migration, this framework allows each edge cloud to dynamically adjust its actions, such as modifying bandwidth or migrating services, in response to the changing needs of the system and its current state.
Figure~\ref{fig:impact_architecture} gives an overview of IMPACT. Each agent first encodes its local history into a latent intent, then combines peer intents and local features through the double-attention module to choose a bandwidth-adjustment or migration action.

\textbf{1) Intent Encoder:} Each agent encodes its trajectory $\tau_i^t=(o_i^1,a_i^1,\dots,o_i^t)$ using a GRU to obtain $h_i^t$. The intent encoder maps $(h_i^t,a_i^{t-1})$ into Gaussian parameters $(\mu_i^t,\delta_i^t)$, and a latent intent variable is sampled:
$
g_i^t \sim \mathcal{N}(\mu_i^t, (\delta_i^t)^2).
$
In the context of edge microservice migration, the intent of each edge cloud represents its short-term goal to either adjust its resource allocation (e.g., bandwidth) or initiate migration of a microservice to balance load and meet service-level objectives (SLOs). Reparameterization $g_i^t=\mu_i^t+\delta_i^t\odot\epsilon, \ \epsilon \sim \mathcal{N}(0,I)$ allows differentiability and controllable stochasticity. The intent encoder thus overcomes the limitation of using only instantaneous observations, encodes historical context, and provides semantic guidance for subsequent attention modules, helping each edge cloud make informed, goal-aligned decisions.

\textbf{2) Message Aggregation:} Intent $g_i^t$ serves as Query, while teammates’ intents $\{g_j^t\}$ act as Key/Value. The attention weights are computed as:
$
\alpha_{ij}^t = \text{softmax}\left(\frac{(W_Q g_i^t)(W_K g_j^t)^\top}{\sqrt{d_k}}\right)$, $
\hat{g}_i^t = \sum_{j\neq i} \alpha_{ij}^t W_V g_j^t + g_i^t,
$
which selects the most relevant teammates, reduces redundant communication, and enhances efficiency by aligning messages with task goals. In the context of microservice migration, the agents (edge clouds) communicate their intent to each other, ensuring that the most relevant edge clouds (those with services that can handle the migration or resource adjustment) are selected to participate in the decision-making process. This minimizes unnecessary communications and optimizes task allocation.

\textbf{3) Environment Attention:} The aggregated intent $\hat{g}_i^t$ is used as Query over local hidden state $h_i^t$:
$
\beta_i^t = \text{softmax}\left(\frac{(W_Q \hat{g}_i^t)(W_K h_i^t)^\top}{\sqrt{d_k}}\right)$,$
\hat{h}_i^t = \sum \beta_i^t W_V h_i^t,
$
which suppresses noisy or irrelevant local features, emphasizes goal-related observations, and improves the interpretability of environmental representation. For edge microservice migration, this allows each edge cloud to focus on relevant environmental information, such as resource usage, service requests, or migration feasibility, while ignoring irrelevant or non-goal-related information (e.g., past states or services already successfully migrated).

\textbf{4) Regularization Losses:} Entropy loss encourages distributional diversity:
$
L_{entropy}=-\frac{1}{2}\sum_{i=1}^{n}\left((\mu_{i}^{t})^{2}+(\sigma_{i}^{t})^{2}-\log((\sigma_{i}^{t})^{2})-1\right),
\label{eq:entropy}
$
while continuity loss enforces temporal smoothness:
$
L_{\text{continuity}} = -\frac{1}{K}\sum_{k=1}^K \cos(g_i^t, g_i^{t-k}),
$
which guarantees discriminability, reduces fluctuations, and enhances the robustness of intent-guided communication. In edge microservice migration, these regularization losses ensure that the intent representations remain meaningful and temporally stable, allowing the edge clouds to adapt efficiently over time to changing network conditions and task demands.

\textbf{5) Total Objective:} The overall training loss is:
\begin{equation}
\mathcal{L}_{\text{total}} = L_{\text{TD}} + \beta_{\text{cont}} L_{\text{Continuity}} + \beta_{\text{ent}} L_{\text{Entropy}},
\end{equation}
which balances task performance with intent quality, enabling semantic-rich and temporally stable intent learning. The loss function guides the agents towards achieving the primary objective of minimizing latency and ensuring successful service migration while controlling for energy and communication overhead. 

In summary, IMPACT establishes a closed-loop pipeline of "intent modeling - communication alignment - environmental filtering - regularization," addressing limitations of static attention. Each module fulfills a distinct role: the intent encoder provides semantic direction, message attention enhances communication efficiency, environment attention improves state relevance, and regularization guarantees robustness. Together, they enable adaptive and goal-aligned communication in dynamic multi-agent tasks, such as edge microservice migration, where efficient resource allocation and timely service migrations are critical for maintaining system performance.

\section{Experiment}
\label{sec:simulation}

\subsection{Experiment Settings}

\footnotetext{We have complete source code and full deployment experiments both in reality and in simulation to assist in reproduction. Various parameters and experimental details are provided in the supplementary materials. Supplementary materials and Github link will be provided upon paper publication due to the double-blind review policy.}

\begin{figure}[h]
    \centering
    \includegraphics[width=\columnwidth]{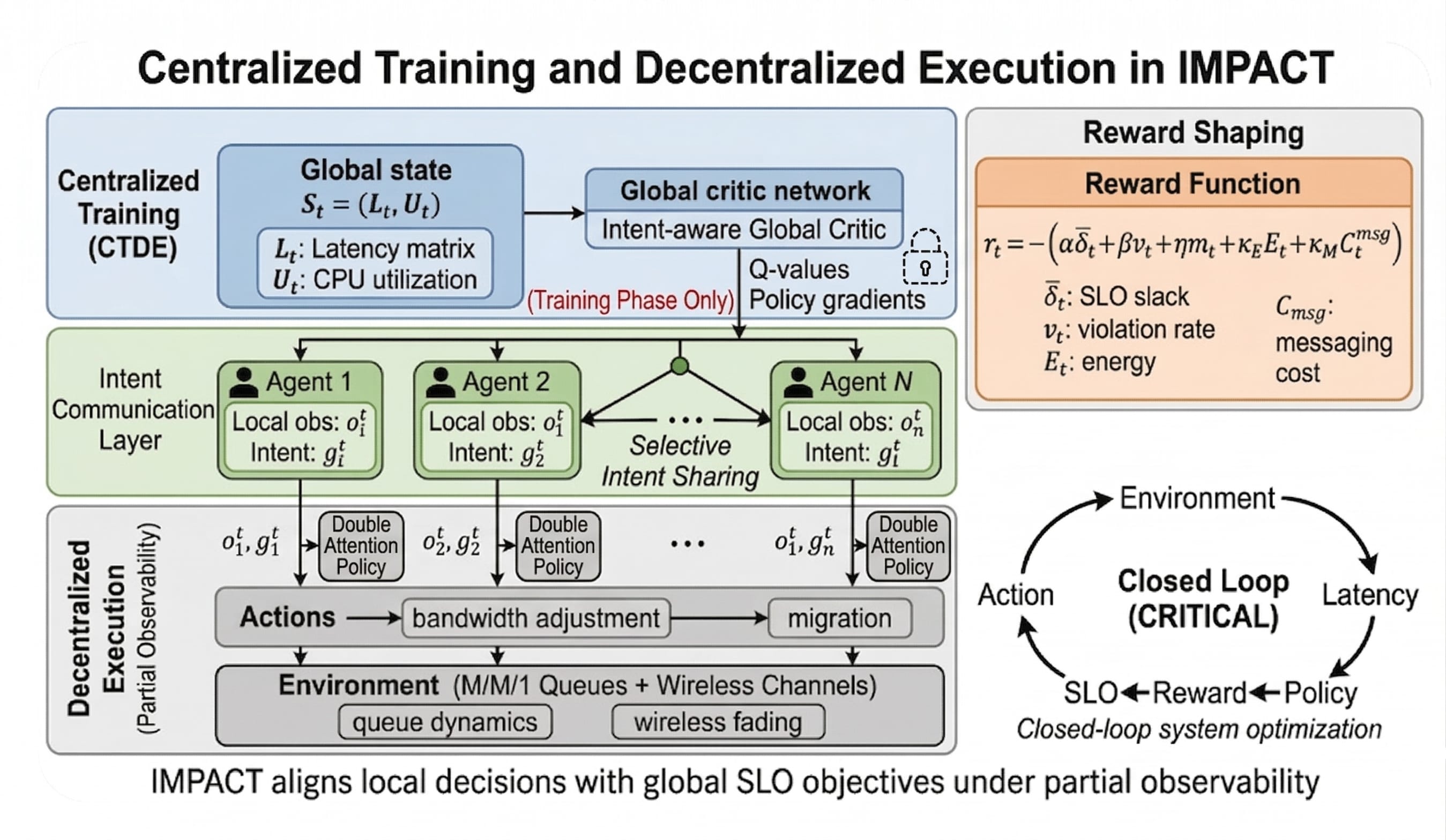}
    \caption{CTDE training and decentralized execution in IMPACT. During training, a centralized global critic provides learning signals using the global state. During execution, each agent acts with local observations, its latent intent, and selectively shared intents, while reward shaping closes the loop with the dynamic environment.}
    \label{fig:impact_ctde_pipeline}
\end{figure}

We evaluate our approach in a custom multi-agent MEC simulator. The system consists of $n{=}20$ mobile users, $m{=}\{5,20\}$ edge clouds, and $k{=}3$ microservice types. We set the latency SLO threshold to $T^{\max}{=}300$\,ms. To demonstrate the effectiveness of our proposed IMPACT, we compare it against six benchmark algorithms: We evaluate our approach in a custom multi-agent MEC simulator. The system consists of $n{=}20$ mobile users, $m{=}\{5,20\}$ edge clouds, and $k{=}3$ microservice types. We set the latency SLO threshold to $T^{\max}{=}300$\,ms. To demonstrate the effectiveness of our proposed IMPACT, we compare it against six benchmark algorithms: Random, Greedy, VDN~\cite{sunehag2018vdn}, QMIX~\cite{rashid2020monotonic}, QPLEX~\cite{wang2021qplex}, and MAIC~\cite{yuan2022multi}.
Fig.~\ref{fig:impact_ctde_pipeline} summarizes the CTDE instantiation of IMPACT. A centralized critic is used only during training, while execution remains decentralized with local observations and selectively shared intents.

\subsection{Experiment Performance}

\subsubsection{Convergence and Network Performance}
\begin{figure}
    \centering
    \includegraphics[width=0.8\linewidth]{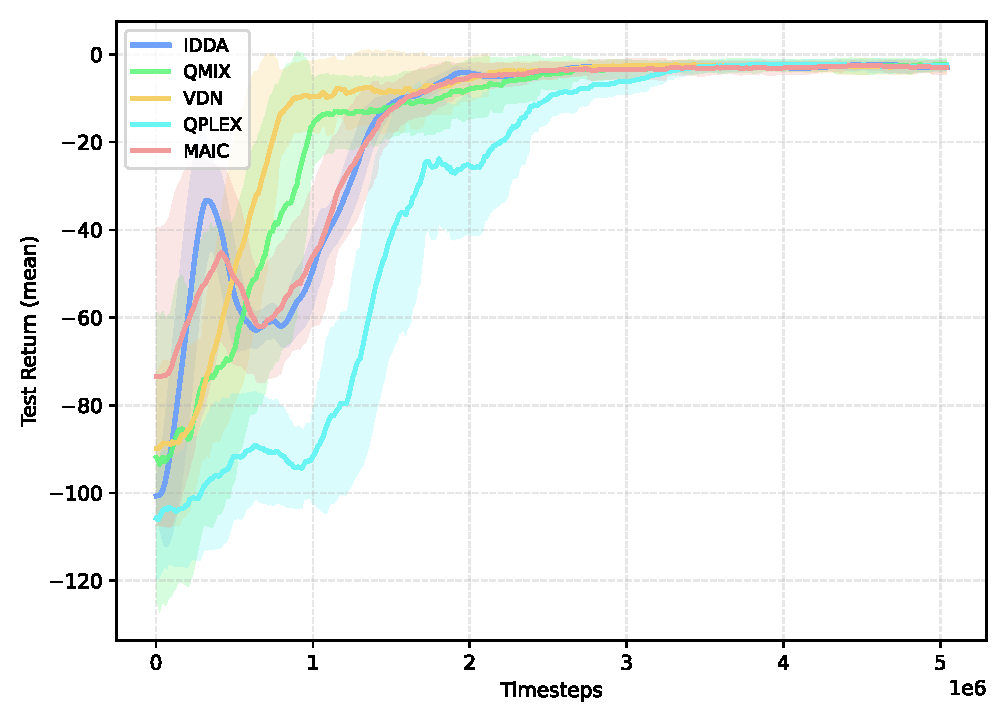}
    \caption{Test return in 20 edges scenario.}
    \label{fig:20 edge return}
\end{figure}

\footnote{We observe a similar trend in other settings (e.g., 5-edge), where IMPACT also performs strongly. Detailed results are in the supplementary and will be include in github.}
When scaling to 20 edges (Table.~\ref{fig:20 edge return}), IMPACT retains its advantage: the curve rises steadily and reaches a high-return plateau with small oscillations. Factorization baselines (QMIX/QPLEX/VDN) converge more slowly with wider confidence bands. QMIX exhibits a prolonged unstable phase before catching up, while VDN and QPLEX converge earlier but still achieve lower final returns. MAIC follows a similar pattern and remains below IMPACT for most timesteps.

In the same setting (Table.~\ref{tab:performance_20_edges}), IMPACT continues to outperform as the network grows. Average latency drops to 93.50\,ms, about 30\% lower than Greedy (133.30\,ms) and 46--50\% lower than VDN (174.70\,ms) and QMIX/QPLEX (187.50/186.50\,ms). Latency deviation is the lowest at 0.047, compared to 0.083--0.091 for Greedy, QMIX, QPLEX, and VDN, and 0.217 for Random, reducing tail variability by 40--50\%. Energy consumption remains competitive at 3.42, close to Greedy (3.40) and below factorized MARL methods (3.67--3.69).

Overall, IMPACT achieves lower mean delay, significantly smaller latency deviation, and comparable or lower energy, showing that intent-driven coordination improves both efficiency and reliability in edge microservice migration. Full results for the 5-edge setting are provided in the supplementary.

\begin{figure}[t]
  \centering
  \includegraphics[width=0.8\linewidth]{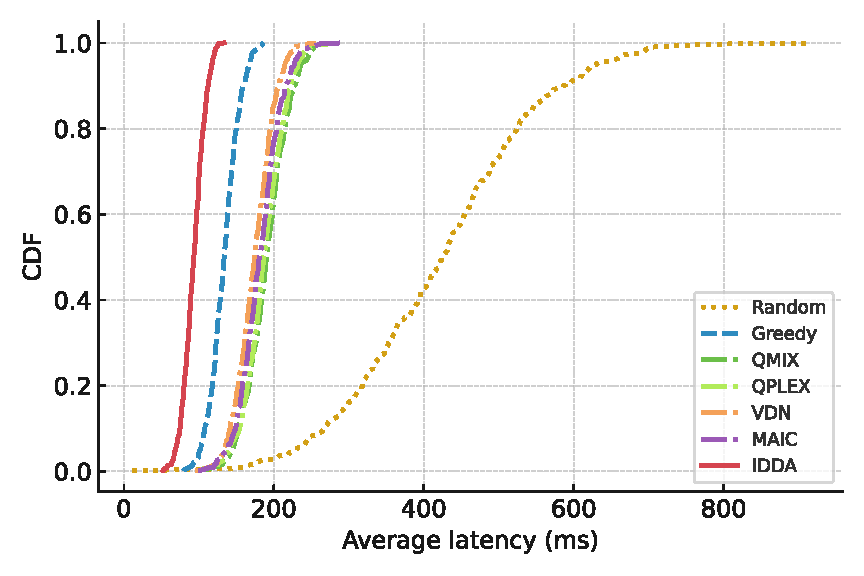}
  \caption{The CDF of service migration average latency in the 20-edge.}
  \label{fig:20 edge cdf}
\end{figure}

\subsubsection{CDF of Average Latency}
We use the cumulative distribution function (CDF) to capture both central tendency and tail behavior under dynamic load and partial observability. It directly reflects SLO percentiles (e.g., P90/P95) and enables visual comparison: a left-shifted, steeper curve indicates lower latency and tighter dispersion.

In the 20-edge case (Fig.~\ref{fig:20 edge cdf}), increased congestion shifts all curves slightly right, but rankings remain unchanged. IMPACT rises steeply, reaching high percentiles at about 120--140\,ms. Greedy follows but requires around 200\,ms or more. QMIX, QPLEX, and VDN cluster behind Greedy with wider spread, while Random shows the heaviest tail and slowest rise. These results confirm that IMPACT reduces both mean latency and dispersion; its steeper curve implies lower variance and fewer high-latency events.

This behavior aligns with the design: the intent encoder compresses SLO risk, migration urgency, and compute pressure into compact messages, and selective sharing avoids broadcast overhead while aligning neighboring actions. Double attention then fuses peer intents with filtered local state, enabling timely migrations before queue buildup and coordinating them with discrete bandwidth adjustments, thereby reducing oscillations from bandwidth inertia and migration cooldown.

\begin{figure}[t]
  \centering
  \includegraphics[width=0.8\linewidth]{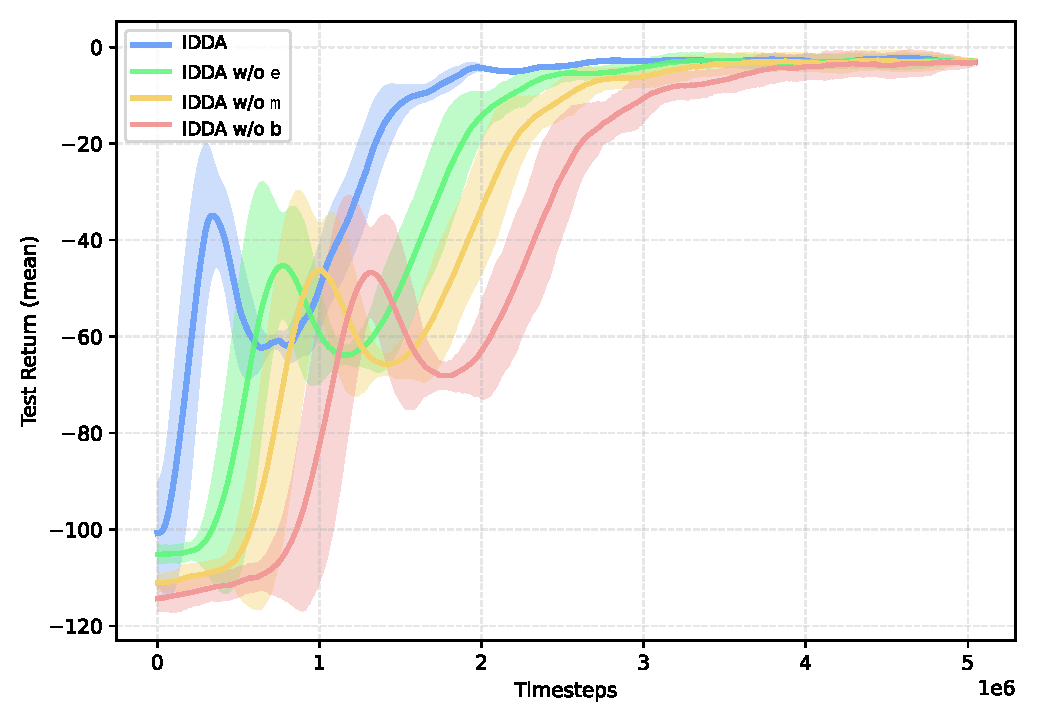}
  \caption{Ablation Result. ``e'' means the loss of environment attention, ``m'' means the loss of message attention, and ``b'' means both losses.}
  \label{fig:ablation}
\end{figure}

\begin{table}[t]
\centering
\caption{Performance comparison on 20 edges. Best results are in bold. Improvements are relative to the best baseline (excluding IDDA).}
\label{tab:performance_20_edges}
\setlength{\tabcolsep}{4pt} 
\renewcommand{\arraystretch}{1.1} 
\begin{tabular}{lccc}
\toprule
\textbf{Method} & \textbf{Cost} $\downarrow$ & \textbf{Latency (ms)} $\downarrow$ & \textbf{Deviation (\%)} $\downarrow$ \\
\midrule
Random  & 3.80 & 423.60 & 0.217 \\
Greedy  & \textbf{3.40} & \textbf{133.30} & 0.088 \\
QMIX    & 3.69 & 187.50 & 0.091 \\
QPLEX   & 3.68 & 186.50 & 0.088 \\
VDN     & 3.67 & 174.70 & \textbf{0.083} \\
MAIC    & 3.66 & 180.90 & 0.087 \\
\midrule
IDDA    & 3.42 {\scriptsize($\uparrow$ -0.6\%)} 
        & 93.50 {\scriptsize($\uparrow$ 29.9\%)} 
        & \textbf{0.047} {\scriptsize($\uparrow$ 43.4\%)} \\
\bottomrule
\end{tabular}
\end{table}

\subsubsection{Ablation Experiments}
Fig.~\ref{fig:ablation} compares full IMPACT with three ablated variants, showing how the two attention components affect learning. The full model rises quickly from negative returns and stabilizes near the optimum with low variance. Removing environment attention (IMPACT w/o~e) slows learning and increases early- and mid-stage oscillations, as unfiltered local features allow transient queue spikes and bandwidth inertia to affect representations. Removing message attention (IMPACT w/o~m) further slows convergence and increases variance: although intents still encode local SLO risk and compute pressure, their unguided fusion weakens inter-edge alignment and leads to mis-timed migrations that shift rather than dissipate hotspots. Removing both (IMPACT w/o~b) yields the slowest, noisiest learning; while it eventually approaches the optimum, the lack of gating and dual attention undermines stable cooperation under partial observability. Although late-stage returns are similar across variants, full IMPACT consistently achieves higher sample efficiency and lower variance, indicating that the two components work synergistically to accelerate convergence and stabilize performance. \footnote{A detailed discussion on system limitations and future plan is available in the supplementary and will be include in github.}

\section{Conclusion}
\label{sec:conclusion}
This paper presents IMPACT, an intent-driven multi-agent communication framework, designed to tackle the challenge of latency-guaranteed microservice migration in mobile edge computing. Unlike traditional MARL and heuristic methods, IMPACT explicitly models agents’ short-term intents, such as SLO risk, migration urgency, and compute pressure, and leverages a double-attention mechanism to adaptively communicate these intents under bandwidth constraints. Our experimental results demonstrate that IMPACT outperforms existing solutions. Compared to QMIX, IMPACT accelerates convergence by over 30\%, reduces P95 latency by up to 25\% in both light- and heavy-load scenarios, and decreases SLO violations by up to 40\%. Furthermore, IMPACT achieves competitive migration costs and energy consumption relative to greedy baselines, while ensuring robust performance in dynamic environments with limited edge resources, bandwidth inertia, and migration cooldown. These results underline the effectiveness of aligning multi-agent communication with decision-making needs. Future work will extend the framework to multi-tenant environments, incorporate advanced migration techniques such as pre-warming, and validate IMPACT on real-world traces and edge prototypes.

\bibliographystyle{IEEEtran}
\ifarxiv
\bibliography{reference,reference_arxiv}
\else
\bibliography{reference}
\fi

\end{document}